\documentclass[twocolumn,english,showpacs,superscriptaddress,prl]{revtex4-1}
\usepackage[latin9]{inputenc}
\usepackage[active]{srcltx}
\usepackage{color}
\usepackage{textcomp}
\usepackage{amstext}
\usepackage{graphicx}
\usepackage{amsmath}
\makeatletter
\DeclareFontEncoding{LGR}{}{}
\DeclareRobustCommand{\greektext}{%
  \fontencoding{LGR}\selectfont\def\encodingdefault{LGR}}
\DeclareRobustCommand{\textgreek}[1]{\leavevmode{\greektext #1}}
\ProvideTextCommand{\~}{LGR}[1]{\char126#1}
\providecommand{\tabularnewline}{\\}

\providecommand{\tabularnewline}{\\}
\usepackage{graphics}\usepackage{subfigure}\usepackage{longtable}\usepackage{pstricks}\usepackage{dcolumn}\usepackage{bm}
\usepackage{babel}
\makeatother
\usepackage{babel}
\begin{document}
\title{Development of the magnetism in the solid solution of the candidate Weyl semimetals Ce$_{1-x}$Pr$_{x}$AlGe}
\author{Pascal Puphal}
\email{pascal.puphal@psi.ch}

\affiliation{Laboratory for Multiscale Materials Experiments, Paul Scherrer Institute,
5232 Villigen, Switzerland}
\author{Sarah Krebber}
\affiliation{Laboratory for Multiscale Materials Experiments, Paul Scherrer Institute,
5232 Villigen, Switzerland}
\author{Emmanuelle Suard}
\affiliation{Institut Laue-Langevin BP 156, 38042 Grenoble Cedex 9, France}
\author{Robert Cubitt}
\affiliation{Institut Laue-Langevin BP 156, 38042 Grenoble Cedex 9, France}
\author{Chennan Wang}
\affiliation{Laboratory for Muon-Spin Spectroscopy, Paul Scherrer Institut, 5232 Villigen, Switzerland}
\author{Tian Shang}
\affiliation{Laboratory for Multiscale Materials Experiments, Paul Scherrer Institute,
5232 Villigen, Switzerland}
\author{Victor Ukleev}
\affiliation{Laboratory for Neutron Scattering and Imaging, Paul Scherrer Institute,
5232 Villigen, Switzerland}
\author{Jonathan S. White}
\affiliation{Laboratory for Neutron Scattering and Imaging, Paul Scherrer Institute,
5232 Villigen, Switzerland}
\author{Ekaterina Pomjakushina}
\affiliation{Laboratory for Multiscale Materials Experiments, Paul Scherrer Institute,
5232 Villigen, Switzerland}
\begin{abstract}
We investigate the macroscopic and microscopic physical properties of the solid solution of Ce$_{1-x}$Pr$_{x}$AlGe. The series tunes from CeAlGe with its multi-$\vec{k}$ structure and a major moment
in the ab-plane, to PrAlGe with an easy-c-axis ferromagnetic ground state
co-existing with a low density of nanoscale textured magnetic domain
walls. Using AC-, DC-susceptiblity, resistivity, specific heat, muon spin relaxation/rotation and neutron scattering we analyze the magnetic ground state of the series. We provide further evidence supporting our previous claim for spin-glass like properties in pure PrAlGe. With introduction of Pr to CeAlGe the finite magnetic field required to stabilize the topological multi-$\vec{k}$ magnetic phase for $x=0$ becomes suppressed. The crossover between the two end-member ground states occurs in the vicinity of $x=0.3$, a region where we further anticipate the field-induced topological magnetic phase for $x < 0.3$ to become the zero field ground state. 
\end{abstract}
\maketitle

\section{Introduction}

Weyl semimetals are a new class of topological conductors where in
the bulk, conduction and valence bands cross at discrete singular
points known as Weyl points \cite{Weyl(1929)}. After the first experimental
evidence for such a state in TaAs in 2015 \cite{Xu(2015),Xu2(2015)}
many new candidates have been explored. Magnetic Weyl semimetal yield the possibility of controlling the Weyl node pattern with an external magnetic field. For this class of materials recent theoretical calculations predicted that the polar magnet family $R$AlGe ($R$ = Ce, Pr) \cite{Chang(2018)} offers remarkable tunability where type- I, type-II inversion-breaking, and time-reversal-breaking types of ferromagnetic Weyl semimetal states are all available. These magnetic non-centrosymmetric semimetals offer rich possibilities for Weyl fermions since both spatial and time-reversal symmetries are broken for which the microscopic states can be tuned via chemical substitution. 

While ternary LaPtSi-type (I4$_{1}$md) $R$AlGe alloys were already discovered in 1992 \cite{Dhar(1992)} the interest of these systems in the context of topological states has only been realized in 2018 \cite{Chang(2018)}. Until recently only structural, specific heat and magnetization data were published on CeAlGe polycrystalline
samples \cite{Dhar(1992),Flandorfer(1998),Dhar(1996)} while for PrAlGe only the crystal structure had been studied previously \cite{Gladys(2000)}. Since the theoretical
proposal of a Weyl state in these materials however, the interest in the two candidates has been recently ignited \cite{Puphal2019,Hodovanets(2018),Suzuki2019,Meng2019}.
We were able to grow stoichiometric single crystals of both CeAlGe ($T_{C} = 5$\,K) and PrAlGe ($T_{C} = 16$\,K) \cite{Puphal2019} by floating zone growth and probe their magnetic structures using small-angle neutron scattering (SANS) and neutron powder diffraction (NPD).
Other reported experimental studies were done on Al-self flux grown crystals which often present an Al enrichment and lead to different physics \cite{Puphal2019,Hodovanets(2018),Suzuki2019}.
While PrAlGe presents the expected easy-c-axis ferromagnetic ground state \cite{Destraz2020},
for CeAlGe we discovered a surprisingly complex multi-$\vec{k}$ structure with a field induced topological magnetic state \cite{Puphal2020}.

Here we present the magnetic ground state evolution tuned by chemical substitution
in polycrystalline materials of the Ce$_{1-x}$Pr$_{x}$AlGe solid
solution. We performed a detailed magnetic characterization, analyzing  AC-, DC-susceptiblity, specific heat, transport, muon spin resonant relaxation/rotation (\textgreek{m}SR) spectroscopy as well as neutron scattering results. As our main result, we find a crossover in the nature of the magnetic groundstate at $x\approx 0.3$, where moreover the magnetic field-stablized topological phase for $x$ slightly above 0.3 likely becomes the zero field ground state for a small region of $x$.

\section{Experimental Details}

The polycrystalline samples were prepared by arc melting with a Compact
Arc Melter MAM-1, the three starting elements Ce/Pr, Al and Ge of
a minimum purity of 99.99\% for homogeneity. 
X-ray fluorescence (XRF) spectra were recorded using the Orbis microXRF analyzer
from EDAX. AC- and DC- magnetic susceptibility were carried
out in a range of  1.8 - 400\,K, 0 - 7\,T and 0.02 - 1480\,Hz, using a Quantum Design
Magnetic Property Measurements System (MPMS). Additionally AC- and DC- magnetic susceptibility measurements, resistivity and specific heat measurements were carried out in the range of 1.8 - 300\,K and 0 - 9\,T, 10-10000\,Hz on a Physical Property Measurement System (PPMS).
Neutron diffraction experiments were done at beamlines at Institute Laue-Langevin (ILL): D11 small-angle neutron scattering (SANS) beamline
in the range of 1.5 - 20\,K with a wavelength of $\lambda=4.6\text{\thinspace\AA}$
using polycrystalline ingots of Ce$_{1-x}$Pr$_{x}$AlGe each of around 300\,mg
glued on an aluminium plate with GEvarnish, D1B at $T=1.5$\,K, 7\,K
and 15\,K with a wavelength of $\lambda=2.525\text{\thinspace\AA}$,
as well as D2B $T=300$\,K using $\lambda=1.594\text{\thinspace\AA}$
both on powderized polycrystalline samples of 2\,g each loaded into
indium sealed vanadium cans in a helium glovebox.
Pressed pellets of powderized polycrystalline ingots of PrAlGe, CeAlGe and Ce$_{0.7}$Pr$_{0.3}$AlGe of 5\,mm diameter were measured on the GPS instrument at the Swiss muon source at PSI, in  zero  magnetic field  (ZF)  and temperatures down to 1.6\,K

\section{Elemental Analysis (EDS)}

A detailed XRF study was performed to analyze carefully the compositions of all polycrystalline materials. This characterization is crucial,
since both the structure-type and physical property transition temperatures
can vary according to small compositional variations of $R$Al$_{x}$Ge$_{2-x}$
\cite{Dhar(1996),Hodovanets(2018)}, which has a tremendous impact
on the physical properties \cite{Puphal2019}. For the XRF study, we used a CeAlGe polycrystalline sample characterized in detail by XRF, energy dispersive spectroscopy (EDS) and neutron diffraction \cite{Puphal2020} as a standard.

\begin{figure}[t]
\begin{centering}
\includegraphics[width=1\columnwidth]{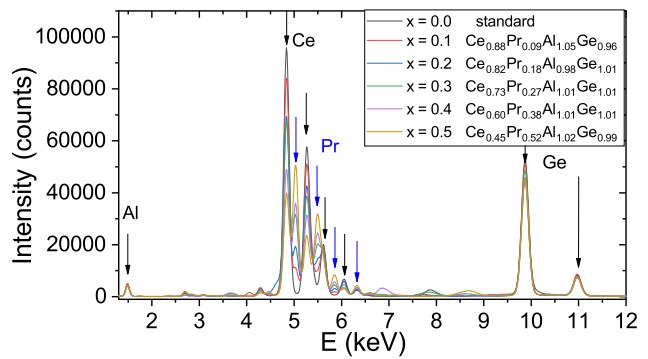}
\par\end{centering}
\caption{\textcolor{black}{\label{EDX} }XRF spectra of the polycrystalline
series Ce$_{1-x}$Pr$_{x}$AlGe measured on polished discs in vacuum normalized
to the stoichiometric CeAlGe polycrystalline sample from Ref. \cite{Puphal2020}.}
\end{figure}

In case of the polycrystalline materials the issues faced for self-flux
growth do not exist and all samples are perfectly homogenous stoichiometric
ingots as determined from the XRF data (see Figure \ref{EDX}). The shown
data are single point measurements with a 30\,\textgreek{m}m wide beam counted for an hour in vacuum.
No averaging was performed, as we saw no variation between the samples
within uncertainty (checked by detailed mapping of a large
polished area). The off-stoichiometry of the determined compositions and listed in the legend of Figure \ref{EDX} is maximally 0.05, and is below the 2 at\% standard error of any elemental analysis. Furthermore this XRF analysis was cross-checked
with an EDS analysis in a scanning electron microscope (SEM) which confirms perfect sample stoichiometries within uncertainty.

\section{Crystal structure}

\begin{table*}
\caption{\label{tab:Crystallographic-data-of}Rietveld refinement results of
the three powder samples of Ce$_{1-x}$Pr$_{x}$AlGe measured at room
temperature with a wavelength of 1.59\, $\textrm{\AA}$ 
on D2B and at 15\,K with a wavelength of 2.52 \AA \,on D1B.}

\centering{}%
\begin{tabular}{l|lll|lll}
T {[}K{]} & \multicolumn{3}{c|}{300} & \multicolumn{3}{c}{15}\tabularnewline
\hline 
x & 0.2 & 0.4 & 0.6 & 0.2 & 0.4 & 0.6\tabularnewline
\hline 
a $[\textrm{\AA]}$ & 4.27884(1) & 4.27449(1) & 4.26957(2) & 4.26003(9) & 4.26290(7) & 4.25638(9)\tabularnewline
c $[\textrm{\AA]}$ & 14.69229(9) & 14.68619(8) & 14.68354(10) & 14.65517 (39) & 14.66125(32) & 14.64836(38)\tabularnewline
$z_{R}$ & 0.59270(20) & 0.59401(18) & 0.59504(19) & 0.59483(38) & 0.59140(29) & 0.59308(35)\tabularnewline
$z_{Al}$ & 0.17626(17) & 0.17795(16) & 0.17896(17) & 0.17432(36) & 0.17127(27) & 0.17738(32)\tabularnewline
$z_{Ge}$ & 0.01001(17) & 0.01131(16) & 0.01240(18) & 0.01319(36) & 0.00642(28) & 0.01246(33)\tabularnewline
R-factor & 3.37 & 3.26 & 3.09 & 4.46 & 3.52 & 3.47\tabularnewline
\end{tabular}
\end{table*}

\begin{figure}
\begin{centering}
\includegraphics[width=1\columnwidth]{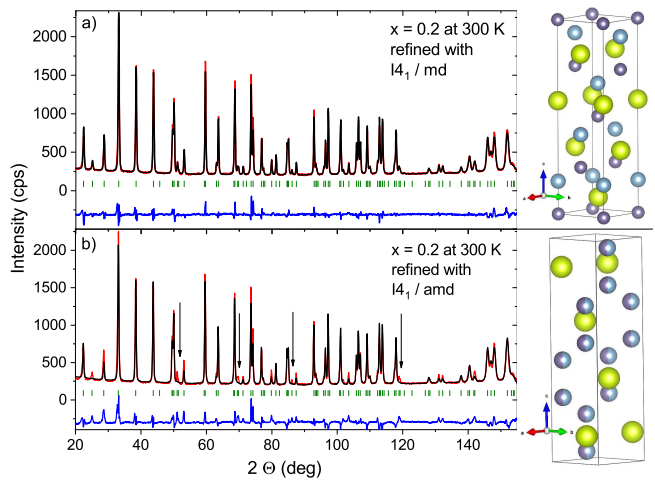} 
\par\end{centering}
\caption{\textcolor{black}{\label{strucdiff} }Rietveld refinement of Ce$_{0.8}$Pr$_{0.2}$AlGe NPD pattern measured at room temperature with a wavelength of 1.59\,\AA  on D2B for two structures reported for CeAlGe a) I4$_{1}$md b) I4$_{1}$amd where arrows highlight the reflexes missing intensity. On the right part the corresponding crystal structure is shown, with Ce in green, Al in blue and Ge in purple.}
\end{figure}

\textit{R}AlGe was first reported to crystallize in the $\alpha-$ThSi$_{2}$
structure-type with space group $I$4$_{1}$/amd (No. 141) \citep{Dhar(1992)},
with later studies instead proposing the LaPtSi structure-type \citep{Dhar(1996),Gladys(2000)} with a body-centered polar tetragonal space-group $I$4$_{1}$md (No. 109). The structures are shown in the right part of Figure \ref{strucdiff}. As apparent from Rietveld refinement results displayed in Figure \ref{strucdiff} both structures result in different intensity distributions in neutron diffraction, where the main difference for the $I$4$_{1}$/amd is highlighted by arrows. From these Rietveld refinements of neutron diffraction data, we find the Ce$_{1-x}$Pr$_{x}$AlGe series forms a solid solution with a stable I4$_{1}$md (No. 109) structure realized for all $x$. The structure is stable down  to low temperatures, and a linear increase of the lattice constant with substitution following Vegard\textquoteright s law (see Table \ref{tab:Crystallographic-data-of} ). Thus we can confirm that for the whole series of Ce$_{1-x}$Pr$_{x}$AlGe with a stoichiometric Al-Ge ratio the LaPtSi-type structure is realized, since powder neutron diffraction allows us to differentiate conclusively between the two candidate structure types. 

\section{Magnetic Properties}

After having proven stoichiometric composition of the samples, we next turn to the detailed characterization of the magnetic properties of Ce$_{1-x}$Pr$_{x}$AlGe in the temperature range from 1.5 to 400\,K. As published recently \cite{Puphal2020},
magnetic Ce ions in CeAlGe order incommensurately below 4.5\,K with a unique multi-$\vec{k}$ ground state; i.e.  spin texture that is characterized by a periodic array of regions for which the magnetic topological winding number is either +1/2 or -1/2, and cancels when averaged over the entire sample. PrAlGe on the other hand orders predomantly ferromagnetically along the c-axis, and presents a small fraction of a co-existing nanoscale-sized domain walls involving moments tilted away from $c-$axis best seen in SANS at low $q$-values. 

\begin{figure}[t]
\begin{centering}
\includegraphics[width=1\columnwidth]{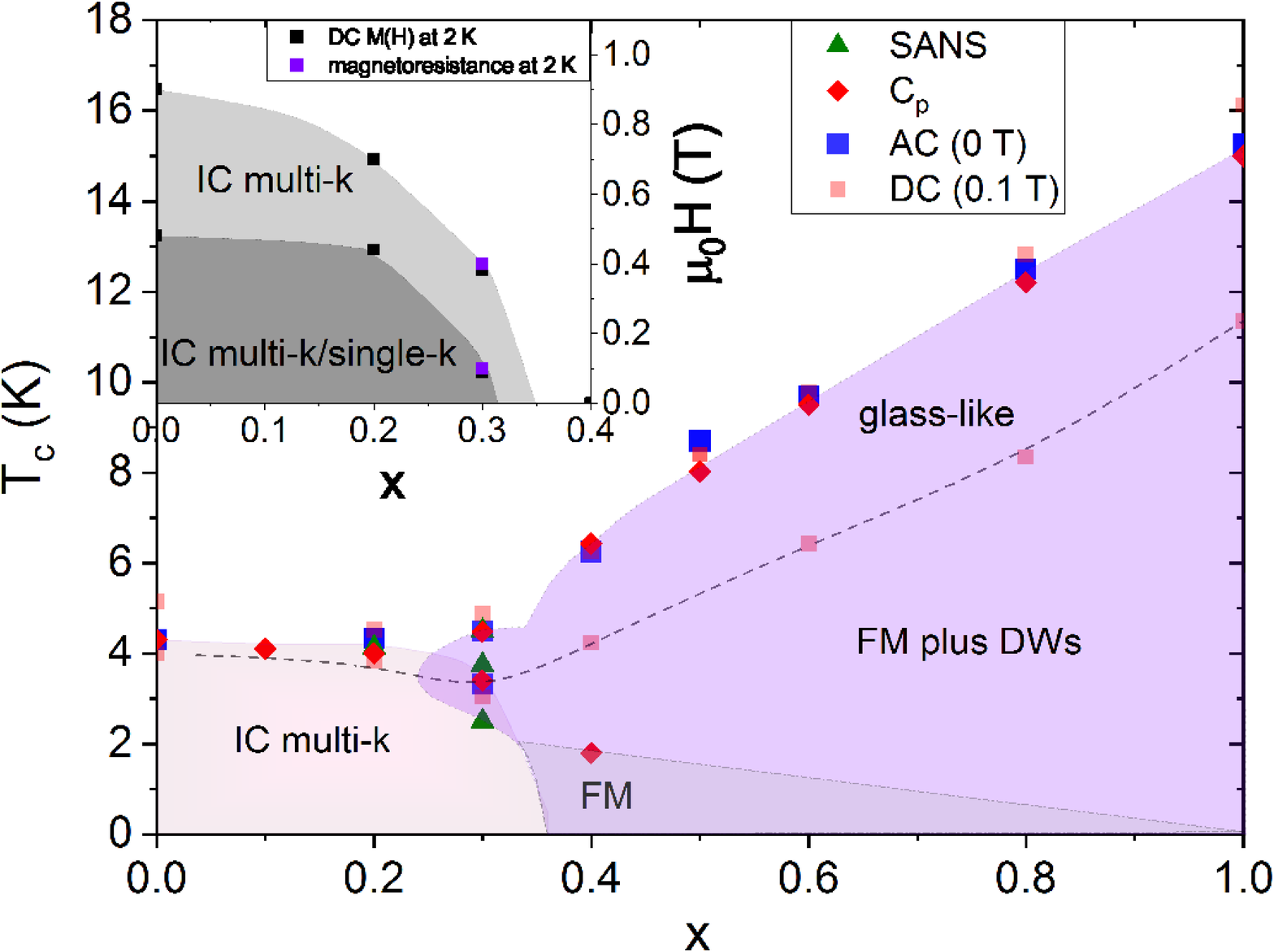} 
\par\end{centering}
\caption{\textcolor{black}{\label{phase} }Resulting phase diagram for
Ce$_{1-x}$Pr$_{x}$AlGe solid solution. The labelled magnetic phases are
IC multi-$\vec{k}$: incommensurate magnetic multi-$\vec{k}$ ground state
determined for CeAlGe \cite{Puphal2020}, FM plus DWs: Predominantly easy-c-axis ferromagnet with low density of nanoscale domain walls, FM: uniform
ferromagnet. The inset shows the field dependence drawn from the derivatives of the $M(H)$ curves shown in figure
\ref{dc derivative}. }
\end{figure}

Figure  \ref{phase} shows the phase diagram for the solid solution series constructed from data presented in this paper. We find a slow change between the end component ground states with a continuous shift of $T_{N}$ resulting in an overlap at a substitution level of $x=0.3$.
With the introduction of Pr to CeAlGe we first find no strong influence
on the ground state, but already a linearly increasing size of the bulk moment (see section V.B). A second transition emerges from Pr spins due to glass-like behavior co-existing with the incommensurately ordered Ce moments, which is suppressed for $x\geq0.3$ as seen in the $T_{N}$ decrease.  With further increase of the Pr content ($x\geq0.4$) all moments, including those due to Ce, align ferromagnetically with the Pr ions with the typical domained deviation \cite{Destraz2020}. From then on the ferromagnetic
ordering temperature and moment increase in a linear fashion with $x$. 
In addition to these ground states we found in SANS for the mixed system in $x=0.3$ a suppression of the low $q$ scattering below 3\,K, indicating a fully aligned ground state, which is discussed in detail in section VI. We find evidences of this transition in $C/T$ drawn from the specific heat as a second maximum in all samples, which decreases in temperature with increasing $x$ (see section V.C).
Further evidence for this transition is given in The field dependence of the multi-$\vec{k}$ ground state is scetched in the inset of Figure \ref{phase}, which predicts a possible stoichiometry, where the magnetic field-induced phase that generates a topological Hall effect signal \cite{Puphal2020} is stable without an external field.

\subsection{AC-susceptibility}

\begin{figure}
\begin{centering}
\includegraphics[width=1\columnwidth]{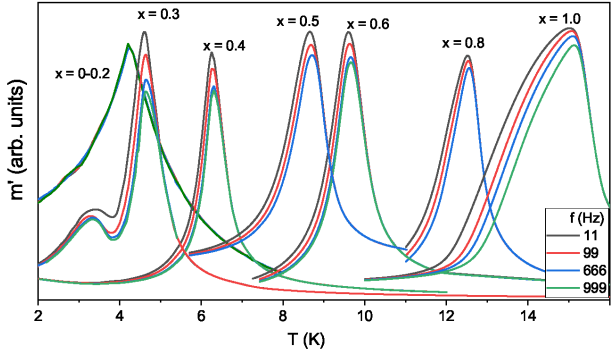} 
\par\end{centering}
\caption{\textcolor{black}{\label{AC freq} AC-susceptibility data at 9, 99, 666 and 999 Hz for the full range of
the substitution series measured in an AC field of 3.9 Oe in the
low temperature range (2-16\,K).}}
\end{figure}

The continuous tuning of the magnetic ground state with $x$ is best seen in the AC-susceptibility data of the full series of the Ce$_{1-x}$Pr$_{x}$AlGe solid solution shown in Figure \ref{AC freq}. Following the maximum of $m'$, we see the decrease of the multi-$\vec{k}$ ground state, just when additionally the ferromagnetic groundstate is realized at $x=0.3$, while the increase of the latter one is apparent starting from $x>0.3$.
While only a subtle frequency dependence is seen in the maximum of $m'$, we observe a stronger shift of the left tail (visible in $m''$ described below). This likely reflects an external frequency-dependent response of the domain dynamics.
The dependence is not apparent for $x=0-0.2$, but starts in the vicinity of both ground states for the $x=0.3$ sample. 
As previously reported also for PrAlGe \cite{Puphal2019} we find for the ferromagnetic ground state a really tiny shift in contrast to the results of Ref. \cite{Meng2019}.

\begin{figure}
\begin{centering}
\includegraphics[width=1\columnwidth]{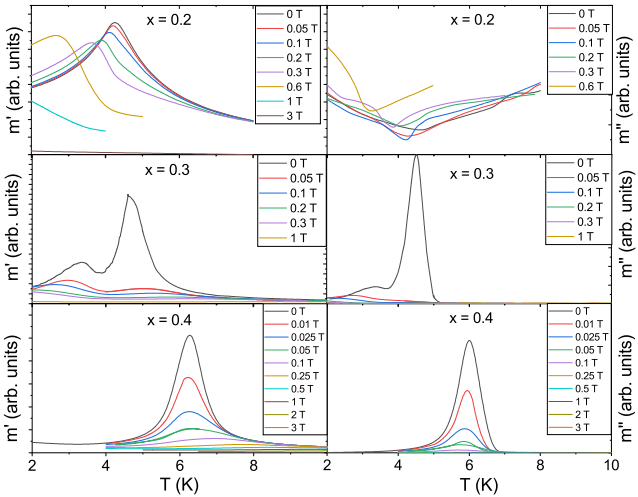} 
\par\end{centering}
\caption{\textcolor{black}{\label{field ac} Field dependence of the AC-susceptibility versus temperature measured at a frequency of 997 Hz, with an AC field of amplitude of 3.8\,Oe for three exemplary compositions $x=0.2-0.4$ measured
around the magnetic ordering transition.}}
\end{figure}

The classification of the underlying ground states is obvious from a difference in the signal strength in $m''(T)$ and in a differing behaviour in external field displayed in Figure \ref{field ac}. The antiferromagnetically coupled incommensurate
ground state for $x<0.4$ shifts down in temperature with increasing
field, while the ferromagnetic ordering transition increases in temperature
with increasing field. In addition larger fields are necessary to
suppress the susceptibility due to the incommensurately ordered phases compared to the ferromagnetic one as shown in Figure \ref{field ac}.

\begin{figure}
\begin{centering}
\includegraphics[width=1\columnwidth]{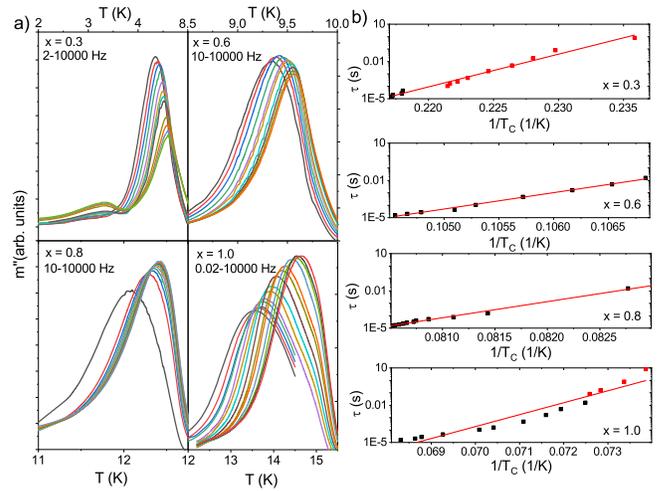} 
\par\end{centering}
\caption{\textcolor{black}{\label{Chi} a) Imaginary part of the AC-susceptibility
measured on the polycrystalline alloys of }Ce$_{1-x}$Pr$_{x}$AlGe
at various frequencies. b) Relaxation time $\tau=1/(2\pi f)$ plotted
versus the inverse temperature of the extracted maxima in a semilogarithmic
scale from AC-susceptibility measurements of $x = 0.3$, 0.6, 0.8 and
1.0. Red squares are from MPMS (QD) measurements, and black squares
from a PPMS (QD). The lines are fits to the Arrhenius law given in
the text.}
\end{figure}

For the magnetic ground state that is predominantly ferromagnetic with the low density of nanoscale domain walls, the AC-susceptibility  shows an aforementioned measurable frequency dependence, which is hardly visible in $m'$, shows up clearly as a shift of the maximum in the imaginary part $m''$. We analyzed the the response of $m''$ from the viewpoint of domain dynamics applying an Arrhenius law $\tau=\tau_{0}\exp(E_{b}/k_{B}T)$ to fit the extracted relaxation
times $\tau$ plotted versus the inverse maxima of $m''$ shown in Figure
\ref{Chi} b) with the linear fit results: $log\tau_{0.3}=(-62(2)+260(10)/T_{C})$~s,
$log\tau_{0.6}=(-142(4)+1320(40)/T_{C})$\,s, $log\tau_{0.8}=(-109(3)+1300(40)/T_{C})$\,s
and $log\tau_{1.0}=(-71(5)+960(80)/T_{C})$\,s shown in Figure \ref{Chi}
b. The resulting energy barriers for $x=0.3$, 0.6, 0.8 and 1.0 are 606\,K, 3033\,K, 2986\,K and 2211\,K with extremely quick relaxation $\tau_{0}\approx$ 1.0$\cdot$10$^{-62}$\,s, 1.4$\cdot$10$^{-143}$\,s, 6.5$\cdot$10$^{-110}$\,s, and 1.1$\cdot$10$^{-71}$\,s.
In the case of $x=0.3$ we find a reduced value indicating an interaction of the two ground states. A clear tendency is apparent once a solely ferromagnetic ground state is realized starting from $x=0.4$, where we find a decrease of the barrier with increasing ferromagnetic contribution of Pr moments. 

\subsection{DC-Susceptibility}

\begin{figure}
\begin{centering}
\includegraphics[width=1\columnwidth]{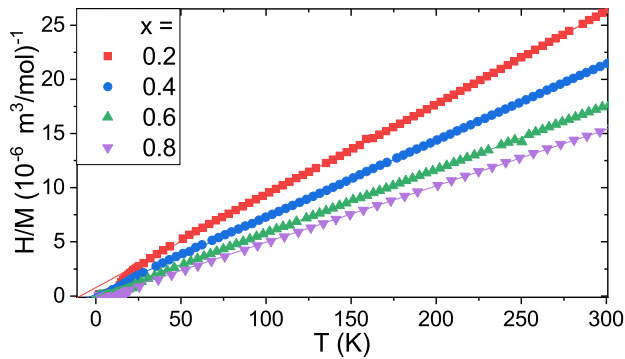} 
\par\end{centering}
\caption{\textcolor{black}{\label{MThigh} Inverse Magnetization versus temperature
in the high temperature range measured at low fields of 0.1 T with applied Curie-Weiss fits.}}
\end{figure}

With the clear image of the transition temperatures drawn from AC-susceptibility we move to the pecularities of the magnetism found in the substitution series analyzed by DC-susceptibility. As both CeAlGe and PrAlGe have a ferromagnetic  $c$-coupling and an antiferromagnetic $ab$-coupling of comparable size seen in the Curie-Weiss fits of data obtained on single crystals \cite{Puphal2019}, for polycrystalline samples one finds a Curie-Weiss temperature close to 0. For CeAlGe with $\Theta_{W}^{||c}=$10\,K and $\Theta_{W}^{||a}=$-42\,K \cite{Puphal2019} ovarious results for polycrystalline materials are reported depending on the orientation and stoichiometry  $\Theta_{W}=-13.5\,K,-25\,K$, -3.5\,K \cite{Dhar(1996),Flandorfer(1998),Hodovanets(2018)}. For PrAlGe we found $\Theta_{W}^{||c}=$36\,K and $\Theta_{W}^{||a}=$-30\,K
compared with values reported elsewhere of $\Theta_{W}^{||c}=$30\,K and $\Theta_{W}^{||a}=$-18\,K \cite{Meng2019}, possibly enhanced by Al-Ge ratio variations. For the polycrystalline samples of the solid solution series we find a reasonable evolution of the Curie-Weiss temperature as one would expect for the slow increase of ferromagnetic contributions
along the $c$-direction. From fits to the data seen in the Figure \ref{MThigh}, we obtain $\Theta_{W}^{0.2}=$-9.1\,K, $\Theta_{W}^{0.4}=-2.55$\,K,
$\Theta_{W}^{0.6}=$ 0.55\,K and $\Theta_{W}^{0.8}=$4.699\,K. With theoretical values for the moment size of Ce of 2.54\,$\mu_{B}$ going to 3.58\,$\mu_{B}$ the effective magnetic moments extracted from fitting high temperature portion of the dc magnetic susceptibility should vary from 2.75, 2.96, 3.16 to 3.37\,$\mu_{B}$ for $x=0.2$, 0.4, 0.6 and 0.8, respectively. Indeed, the effective moment obtained from the Curie-Weiss fits yield: $\mu_{eff}^{0.2}2.73\,\mu_{B}$, $\mu_{eff}^{0.4}2.99\,\mu_{B}$, $\mu_{eff}^{0.6}3.09\,\mu_{B}$ and $\mu_{eff}^{0.8}3.3\,\mu_{B}$.

\begin{figure}
\begin{centering}
\includegraphics[width=1\columnwidth]{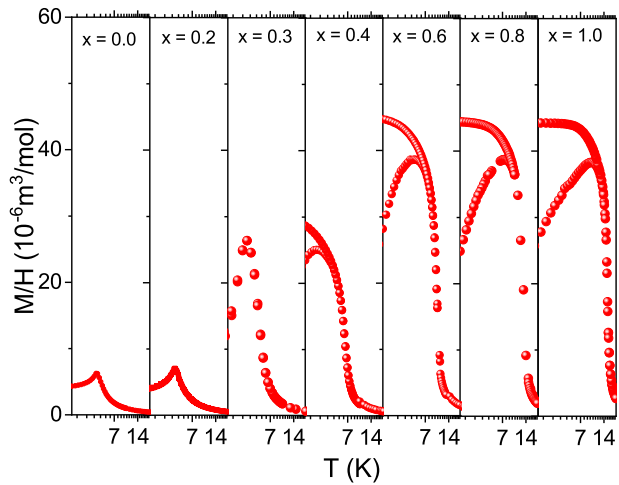} 
\par\end{centering}
\caption{\textcolor{black}{\label{susc Ce} Magnetization versus temperature
in the low temperature range (2-20\,K) measured at low fields of
0.1 T both in field-cooled (upper curve) and zero field-cooled (lower curve) manner.}}
\end{figure}

As seen in Figure \ref{susc Ce}, the low temperature DC-susceptibility shows the expected trend
of increasing magnetization and ordering temperature while adding
more Pr. Again the behavior can be divided into two kinds: while $x<0.4$
shows a downturn in field cooling (FC), we find, as expected for a
ferromagnetic ground state, a saturation to a finite magnetization starting
from $x\geq0.4$. For all $x$, the zero field-cooled (ZFC) measurement shows an increase of the magnetization with temperature up to the ordering temperature. While
pure CeAlGe and low $x$ concentrations and the low substitution of Pr presents a tiny hysteresis at 2\,K \cite{Puphal2019}, for the polycrystalline $x=0.3$ sample
we find no gap between FC and ZFC measurements, despite the significant frequency-dependence of the ac susceptibility indicating domain dynamics as shown in the previous subchapter.

\begin{figure}[h]
\begin{centering}
\includegraphics[width=1\columnwidth]{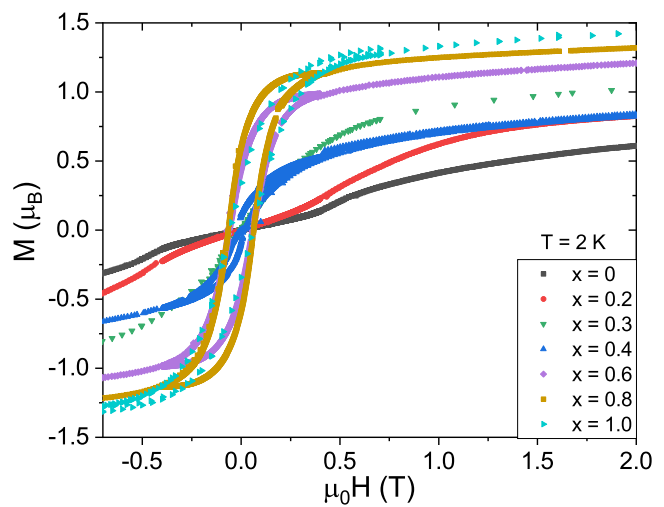} 
\par\end{centering}
\caption{\textcolor{black}{\label{susc Ce-1} Field dependence of the magnetization in the range -0.5 to 2\,T of the magnetization measured on }examplous polycrystalline
samples of the Ce$_{1-x}$Pr$_{x}$AlGe series.}
\end{figure}

Before analyzing the transitions in more detail we look into the hysteretic
behavior seen in field-dependent DC-susceptibility. In Fig. \ref{susc Ce-1}
the magnetization measured at 2\,K for field sweeps is shown in between
-0.7 and 2\,T. We find as expected an increase in the saturation magnetization
 at high fields that scales with the Pr amount. In the $M(H)$ curve the underlying ground state is easily extracted by the fact that once the FM with DWs is realized a hysteresis is observed. In agreement with the
AC data we find a sudden change in the magnetization behavior with
a substitution amount from $x=0.4$, as starting from $x\geq0.4$ the hysteresis
emerges with a similiar coercive field for $x\geq0.6$ of roughly 0.064\,T.

\begin{figure}[h]
\begin{centering}
\includegraphics[width=1\columnwidth]{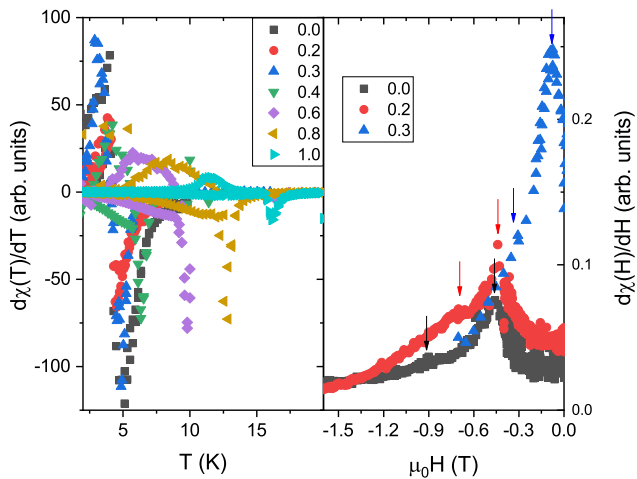} 
\par\end{centering}
\caption{\textcolor{black}{\label{dc derivative} Temperature-derivative (left panel)
and field-derivative (right panel) of the dc susceptibility, revealing $x$-dependence of the transition temperatures
and fields. Arrows highlight maxima in the field-derivative of the DC-susceptibility.}}
\end{figure}

Derivatives of the DC-susceptibility data as shown in Figure \ref{dc derivative}
enables both a visualization and determination of the critical transitions for the temperature, as well as the fields which are added to the phase diagram in Figure
\ref{phase}. In the temperature derivative we always find a positive
and a negative maximum, which for $x=1$ can be accounted to a spin
glass like transition and a subsequent spin reorientation as shown
in Ref. \cite{Puphal2019}. This reorientation transition is only visible in small fields (here 50 Oe) and can be tracked for all substitution amounts even in pure CeAlGe. These data amount to the dashed line in the phase diagram (see Figure \ref{phase}).
From the extrapolation of the critical magentic fields shown in Figure \ref{dc derivative} b) we anticipate a stable magnetic topological state at zero external magnetic field for a substitution of around 33\% Pr.

\subsection{Specific heat}

\begin{figure}[h]
\begin{centering}
\includegraphics[width=1\columnwidth]{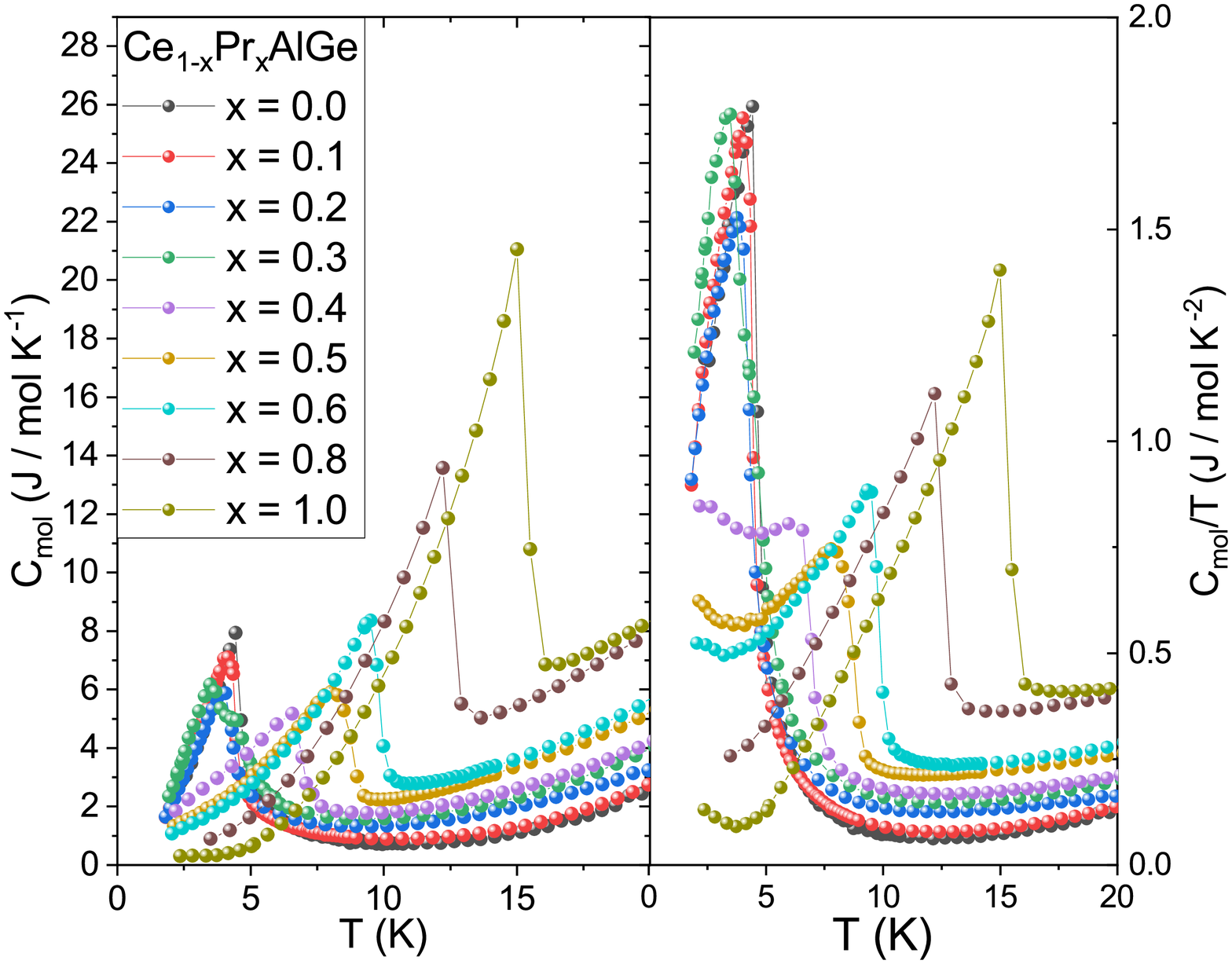} 
\par\end{centering}
\caption{\textcolor{black}{\label{HC} Specific heat of the solid solution
series Ce$_{1-x}$Pr$_{x}$AlGe for the low Pr substituent in the
range of 1.8 - 20\,K in zero field. The right panel shows the same selected
samples of the specific heat divided by temperature. }}
\end{figure}

From specific heat studies of the polycrystalline series we can identify
two transitions. The spin reorientation seen in the derivative of the DC-susceptibility is not observed in specific heat, as was already reported for PrAlGe \cite{Meng2019}. However an additional signal at low temperatures
is found from the vanishing of the nanoscale domain wall texture that we describe in
the next section. The entropy released at the transition is for all cases
around $S\approx R$ln2, when extrapolating the low temperature part of the specific heat for low $x$ concentrations. However the entropy distribution varies with $x$ as for PrAlGe the transition is less sharp. CeAlGe is a $S = 1/2$ system thus a transition of $S\approx R$ln2 describes a full order. The observation of a full order in CeAlGe is in no contrast to the reduced moment
observed in magnetic measurements, as we could prove with XPS that for CeAlGe we find a mixed valence of roughly 50\% Ce$^{4+}$, which is nonmagnetic (see supplemental information of Ref. \cite{Puphal2020}).  For PrAlGe with $S = 1$ an entropy of $S\approx R$ln2  accounts only to 63\%. The missing entropy is in god agreement with the obtained lowered moment (see chapter VI neutron diffraction) supporting our idea of a glassy transition. The reduced moment of 2.29(3)\,$\mu_{B}$ compared to a full Pr$^{3+}$ moment of 3.58\,$\mu_{B}$ resides to 64\%. However part of the missing entropy might lie in the second maximum at lowest temperatures, which is already apparent by the upturn at lowest temperatures shown in the right panel of Figure \ref{HC}. The temperature of the second maximum of $x=0.3$ exactly matches the temperature of the disappearance of low q scattering seen in SANS (see section VI) and indicates a possible spin reorientation to a fully aligned case without domain walls, shown as a grey area in the phase diagram (see Figure \ref{phase}).

\section{Neutron diffraction}

\begin{figure}[h]
\begin{centering}
\includegraphics[width=1\columnwidth]{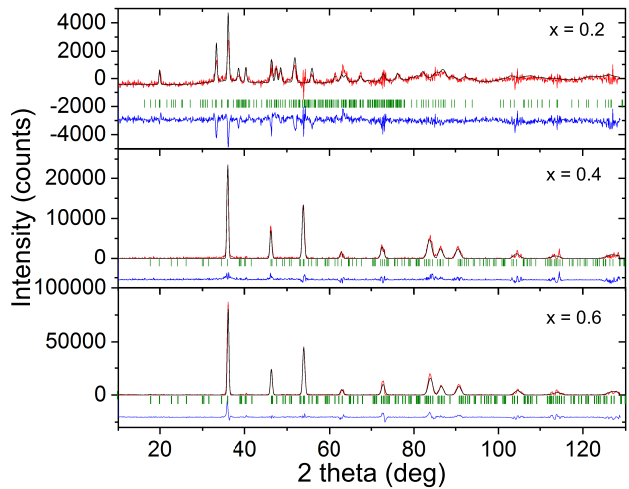} 
\par\end{centering}
\caption{\textcolor{black}{\label{D1B}}NPD difference profiles obtained at
D1B from Ce$_{0.8}$Pr$_{0.2}$AlGe, Ce$_{0.6}$Pr$_{0.4}$AlGe and
Ce$_{0.4}$Pr$_{0.6}$AlGe at base $T = 1.6$\,K below $T_{C}$, and at
15 K above $T_{C}$. The refinement of the magnetic profile is given
in black. The blue line shows the difference between the refined model
profile and the data. The row of green ticks denote possible positions
for magnetic peaks according to the magnetic structure model. }
\end{figure}

In Fig. \ref{D1B}, we show the magnetic contribution, i.e. the difference
pattern of the neutron powder diffraction data measured below (1.5\,K)
and above (15\,K) $T_{C}$ with a wavelength of $\lambda=2.525\text{\thinspace\AA}$,
of three compositions of $x=0.2$, 0.4, 0.6. Similarly as for results obtained from bulk measurements, we find a sharp change of the ground state
above x \textgreater{} 0.3, where the Ce moments presumably become aligned
with the ferromagnetically ordered Pr ones. In contrast, for a low Pr substitution
amount the incommensurate order of pure CeAlGe prevails. To determine the
IC magnetic structure in Ce$_{0.2}$Pr$_{0.8}$AlGe we used the Fullprof
suite \cite{Rodriguez-Carvajal1993} to perform a detailed Rietveld
refinement of the neutron powder diffraction (NPD) difference pattern
against several symmetry-allowed models. The symmetry analysis shows
that the magnetic structure model of highest symmetry is based on
the full propagation star of $\vec{k}$ which consists of four arms
$\vec{k_{1}}=\text{\textpm}(a,0,0)$ and $\vec{k_{1}}=\text{\textpm}(0,a,0)$.
This corresponds to a multi-$\vec{k}$ model described by the maximal
symmetry superspace group I4$_{1}$md1\textquoteright (a00)000s(0a0)0s0s
\cite{Puphal2020}, and has just four refinable magnetic mode amplitude
parameters. Via lebail fitting we find for $x = 0.2$ a slightly larger $\vec{k}$ vector of $a=0.0785(1)$ compared to $a=0.071(1)$ in the case $x = 0$. For $x>0.3$ we find magnetic scattering only at scattering angles commensurate with the tetragonal point symmetry of the \emph{R}AlGe lattice, which can be described by a propagation vector $Q=0$. As for PrAlGe \cite{Destraz2020} for the three symmetry-allowed magnetic structure models for $Q = 0$, we find the standard irreducible representation \textgreek{G}1, that describes a ferromagnetic (FM) order with moments aligned with the c-axis, to solve our magnetic pattern for both $x$ = 0.4 and $x$ = 0.6. The size of the ferromagnetic
moment refined from the data increases as expected with the larger Pr content 1.17( 1)\,$\mu_{B}$ ($x$ = 0.4) and 1.770(1)\,$\mu_{B}$
($x$ = 0.6) compared to 2.29(3)\,$\mu_{B}$ for $x=1$ \cite{Destraz2020}.

\begin{figure*}[t]
\begin{centering}
\includegraphics[width=2\columnwidth]{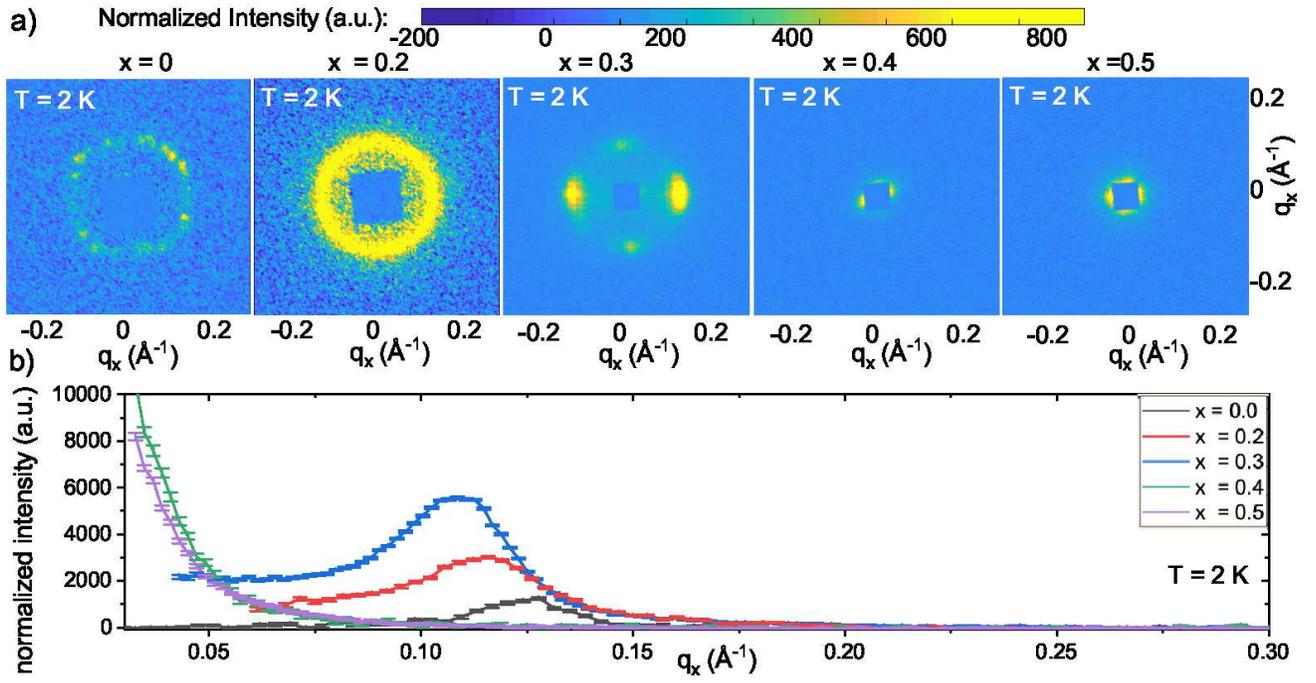} 
\par\end{centering}
\caption{\textcolor{black}{\label{SANS}a) SANS magnetic scattering patterns for five polycrystalline ingots, where the intensity scaling is fixed on the same range for comparison. b) Integrated intensities
of the SANS patterns plotted versus the wave vector $q$ normalized to
the sample size so that intensities can be compared.}}
\end{figure*}

To track the change of the microsopic magnetism with $x$ in more
detail we performed SANS measurements
on polycrystalline ingots of the Ce$_{1-x}$Pr$_{x}$AlGe series. The SANS
images of reciprocal space clearly reveal a sharp change in the general form of low $q$ magnetic scattering above $x>0.3$. For $x < 0.3$, scattering is observed at finite wavevector in analogy with that seen for pure CeAlGe \cite{Puphal2020}, and in the present polycrystallines samples displays either a ring- or Bragg spot-like distribution. This scattering at finite wavevector disappears for x > 0.3, and instead we observe significant magnetic SANS to  emerge  at lowest $q$ for $T < T_{C}$ similar  as for PrAlGe following the phase diagram depicted in Figure \ref{phase}. This observation suggests that in addition to dominant easy-c-axis ferromagnetic correlations, a co-existing nanoscale magnetic texture also exists in the samples of Ce$_{1-x}$Pr$_{x}$AlGe for $x\geq0.3$ analogously as for $x=1$ and discussed in detail in Ref. \cite{Destraz2020}.
In this case the intensity is distributed uniformly in azimuth around the origin,
and falls monotonically over an extended range of \textbar $q$\textbar .

\begin{figure}[h]
\begin{centering}
\includegraphics[width=1\columnwidth]{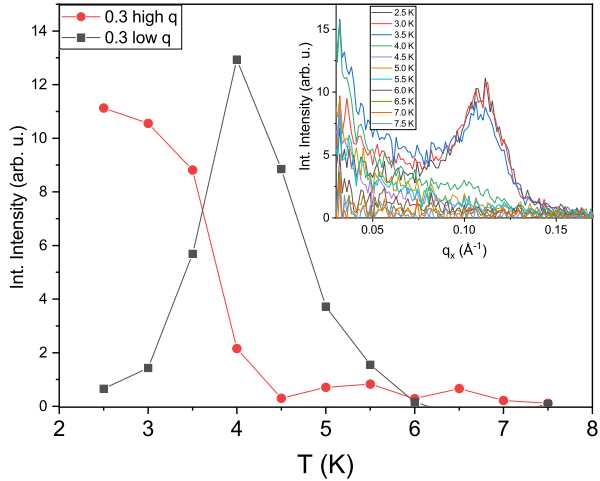}
\par\end{centering}
\caption{\textcolor{black}{\label{SANS-1} Integrated intensities of the SANS
patterns of of $x = 0.3$ plotted versus temperature. On the Inset the $q$ dependent Integrated Intensities at various temperatures is shown.}}
\end{figure}

In addition we find a low $q$ scattering from the slightly misaligned domains already for $x=0.2$ and 0.3 meaning that with the introduction of Pr to the structure, the ferromagnetic coupling part creates a separation in domains (see
Figure \ref{SANS} b). From a proper normalization of the scattered intensity, we can
generally track the increase of scattering contribution due to moment increase
from Pr introduction, though we note that may be compensated by the different intensity distributions observed for $x=0$,0.2 (ring-like) that differ markedly from the almost Bragg-spot like distribution observed from the $x=0.3$ sample. By plotting the same integrated intensity versus
temperature shown in Figure \ref{SANS-1}, we find that the low $q$
scattering vanishes below a certain temperature meaning that the Ce
moments finally align with the Pr ones.

\section{Muon spin relaxation/rotation spectroscopy}

\begin{figure}[h]
\begin{centering}
\includegraphics[width=1\columnwidth]{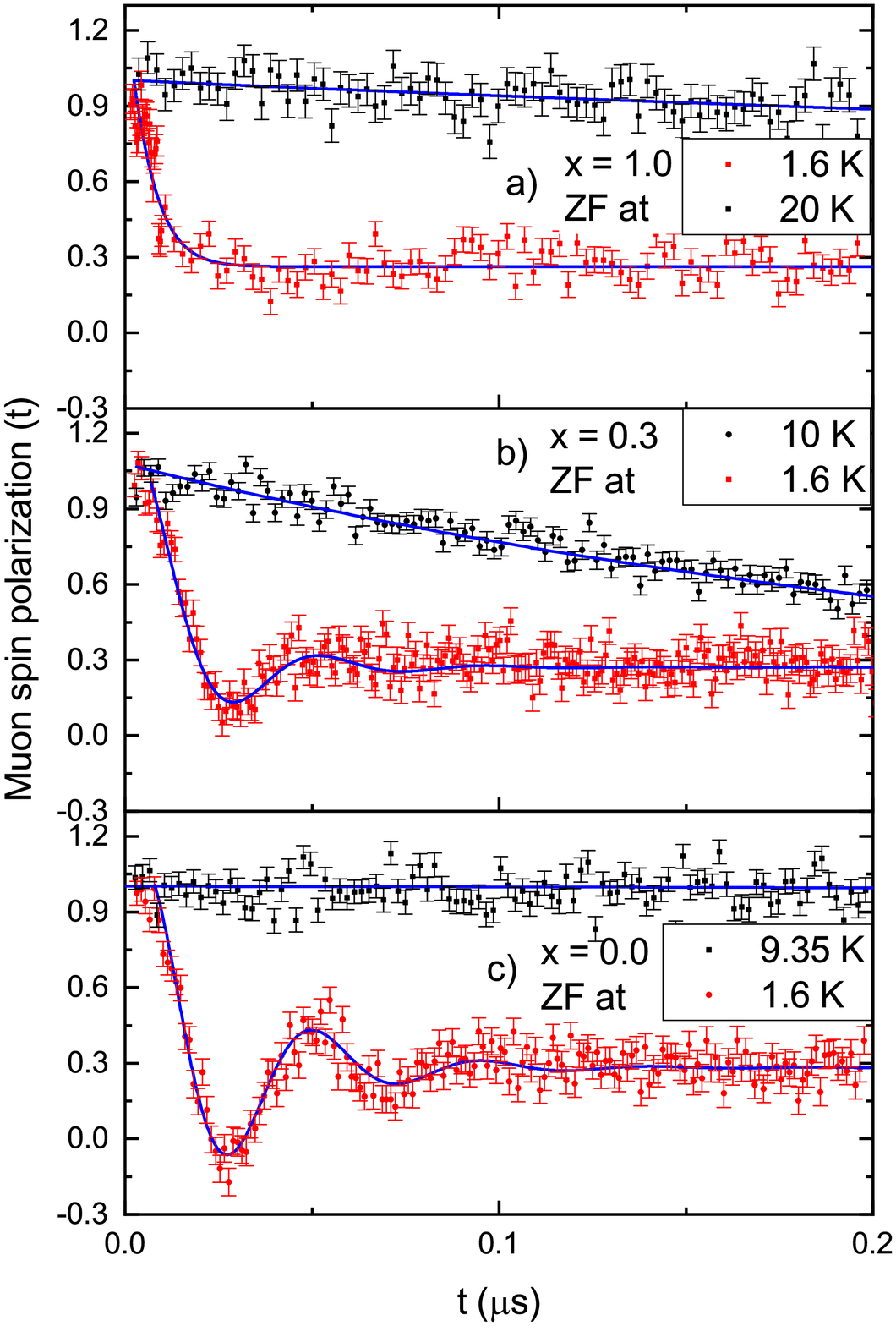}
\par\end{centering}
\caption{\textcolor{black}{\label{musr} Temperature evolution of the muon
spin polarization in the a) $x = 1$, b) $x = 0.3$ and c) $x = 0$ compound in zero-field, at base temperature 1.6 K and far above the transition, i.e. 9.35\,K, 10\,K, 25.2\,K. The solid lines are fits according to Eq. (1) and (2).}}
\end{figure}

To further explore the low-temperature magnetic properties of the Ce$_{1-x}$Pr$_{x}$AlGe series, we conducted zero field (ZF) \textgreek{m}SR experiment down to 1.6\,K. Almost 100\% spin polarized surface muons carrying positive charges are implanted into the sample where they preferentially stop at the most electronegative location near Ge at time zero. The implanted muons relax via coupling with the local magnetic field at the muon site. The temporal evolution of the muon spin polarization is thereafter measured by the asymmetry of the decayed positrons which are emitted preferentially along the muon spin direction \cite{Blundell1999}.
The muon time spectra of the $x = 1.0$ sample measured at 1.6\,K shown in Figure \ref{musr} a) reveals two distinct relaxation signals. A fast decaying component at lowest temperatures showing no oscillatory signal suggests an incoherent magnetic structure developed in the sample, and indicates a freezing of an inhomogeneous distribution of internal fields that supports the existence of a glassy magnetic regime at 1.6\,K \cite{Puphal2019, Destraz2020}. The time-spectrum can be well described by a stretched exponential \begin{equation}P=\frac{1}{3}+\frac{2}{3}\exp(-(\lambda \cdot t)^{\beta})\end{equation} where we obtain a $\beta=1.0(3)$ at 1.6\,K.
 The second long term relaxation can equally well be described by a stretched exponential revealing a maximal beta around the 15\,K transition and a decrease for lower temperatures with $\beta=0.70(2)$ at 14.7\,K reducing down to 0.048(5) at 1.6\,K.
In case of CeAlGe ($x = 0$) (see Figure \ref{musr} c), the asymmetry plot shows a clear damped signal oscillating around 1/3 of the initial spin polarization. The data suggests the muon ensemble detects a coherent internal field induced by the ordered magnetic structure, which is present in the entire volume of the sample. Incommensurate magnetic orders such as spin density waves are best fit by a zeroth order Bessel function $J_{0}$ \cite{Sugiyama2006}. On the other hand, the polarization in the present case of a multi-$\vec{k}$ incommensurate magnetic structure can be described equally well by an exponentially decaying cosine:
\begin{equation}P=\frac{1}{3}+\frac{2}{3}\exp(-\lambda t)\cdot cos(\omega t)
\end{equation}
The maximum frequency is centered at 22.3(3)\,MHz, while the one for PrAlGe is centered at 0 typical for a SG system. 
For the crossover stoichiometry of Ce$_{0.7}$Pr$_{0.3}$AlGe (see Figure \ref{musr} b) we still observe oscillations, which are more strongly damped than for CeAlGe due to the additional PrAlGe character. The maximum frequency is 22(1)\,MHz and hence nearly unchanged compared to the $x$ = 0 case. This suggests that incommensurate multi-k structures observed for x=0 and x=0.3 can be considered to be broadly similar, which is in agreement with data obtained from neutron scattering. 
From the muon data, we find generally large values of the relaxation rate for $R$AlGe. Due to the fast relaxation from the glassiness induced by Pr we find an increasing relaxation rate \textgreek{l} with increasing amount of Pr, which is a hallmark of the broadening of the distribution of the local magnetic fields. In the ordered phase at 1.6\,K we find \textgreek{l}  values of the fast decaying part for increasing $x$ of: 37(1)\,\textgreek{m}s$^{-1}$, 61(3)\,\textgreek{m}s$^{-1}$ and finally 160(20)\,\textgreek{m}s$^{-1}$ for $x$ = 0.0, 0.3, 1.0. 


\section{Crossover region around $x = 0.3$}

\begin{figure}[h]
\begin{centering}
\includegraphics[width=1\columnwidth]{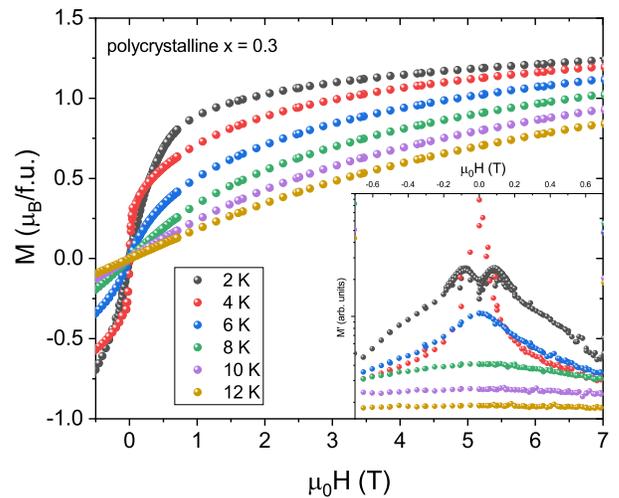}
\par\end{centering}
\caption{\textcolor{black}{\label{0p3-M} Field-dependence of the magnetization at various temperatures in the range of 0-7\,T measured on} a 28.4 mg
ingot of the polycrystalline Ce$_{0.7}$Pr$_{0.3}$AlGe sample.}
\end{figure}

Around Ce$_{0.7}$Pr$_{0.3}$AlGe we find the most complex part in the phase diagram,
as we find an overlap of low $q$ scattering arising from ferromagnetic domains and the multi$-\vec{k}$ scattering around 0.11\,$\AA$ between 3 and 4\,K as well as seperated parts of each as shown in Fig. \ref{SANS-1} e). Indicating seperate regions of the two ground states. We focus now on the details of this compositional area by analyzing the field dependent magnetization and
resistivity. As one might deduce from the indication of frequency-dependent domain dynamics by
AC-susceptibility at 4\,K, where we are in the low $q$ scattering maximum seen in  Figure
\ref{SANS-1}, we find a tiny hysteresis for $x$ = 0.3 seen in the red curve of Figure \ref{0p3-M}. This hysteretic behaviour however vanishes when the sample is cooled down to 2\,K, and for which the low $q$ scattering disappears and we only realize the multi-$\vec{k}$ ground state.

\begin{figure}[h]
\begin{centering}
\includegraphics[width=1\columnwidth]{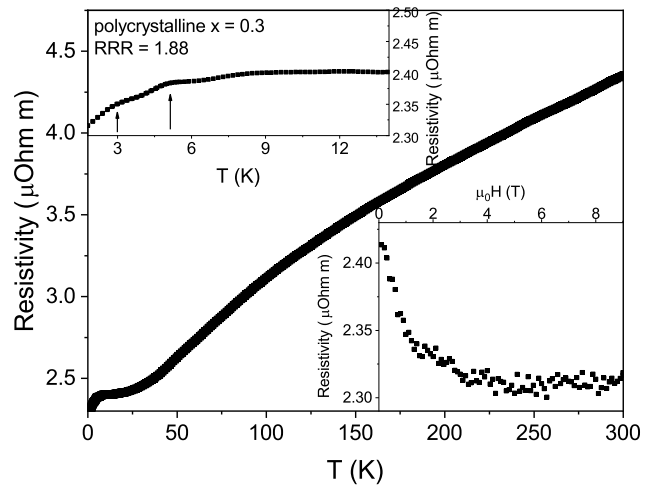}
\par\end{centering}
\caption{\textcolor{black}{\label{0p3} Temperature dependence of the resistivity
on a polycrystalline ingot polished to a flat needle. The inset shows
the field dependent resistivity} measured at 2\,K.}
\end{figure}

For the $R$AlGe samples a relatively low residual resistivity ratio
(RRR) of 1-2 is reported, considering it is a semimetal \cite{Puphal2019,Hodovanets(2018),Meng2019}. So, unsurprisingly we find a decrease of the resistivity with decreasing temperature of the polycrystalline Ce$_{0.7}$Pr$_{0.3}$AlGe sample with a RRR value of 1.88 shown in Figure \ref{0p3}. Similar to the end-compounds we see the magnetic
transition displayed on the inset, which in this case even indicates a double peak with maxima both at 2.8 and 5\,K in good agreement with anomalies observed in AC-susceptibility and SANS measurements (see Figure \ref{SANS-1}). Similar to CeAlGe we find a negative magnetoresistance, the metamagnetic transition is seen in two kinks highlighted by arrows at 0.1\,T and 0.4\,T. In pure CeAlGe, these two metamagnetic transitions occur at higher fields of 0.5 \, and 0.8\,T respectively (see Figure \ref{phase} inset), and over this range of magnetic field a topological magnetic phase that generates a topological Hall signal is stabilized \cite{Puphal2020}. The present study shows that the two metamagnetic transition fields become suppressed with increasing $x$. We thus anticipate that for a further small increase of $x$ beyond 0.3, the field-induced topological phase for $x=0$ can be stabilized as a zero-field ground state.

\section{Summary}

The solid solution Ce$_{1-x}$Pr$_{x}$AlGe preserves the two ground states of the
parent compounds up to the overlapping region around Ce$_{0.7}$Pr$_{0.3}$AlGe,
which represents the general crossover composition between the end member ground states. While for $x<0.3$ the multi-$\vec{k}$ ground state of CeAlGe remains nearly unchanged in its magnetic ordering temperature the metamagnetic transition fields shift leading to the possibility to stabilize the field-induced topological phase as a zero-field ground state when $x$ is just above 0.3. The
predominantly ferromagnetic ground state of PrAlGe decreases in $T_{C}$ with increasing amount of
Ce, so does the moment size, while the energy barrier of the domain formation increases,
which enables a fully FM aligned $\Gamma$- point solution at lowest
temperatures for the mixed system. With Ce$_{0.7}$Pr$_{0.3}$AlGe
we find an interesting compound where the interplay between these
ground states can be studied, as the critical temperatures overlap
only in a limited range and the possiblity of a zero field topological phase is at hand. It provides a candidate Weyl semimetal that enables a tuning of the bands via a purely multi-$\vec{k}$ order below 3\,K and a solely ferromagnetic ground state from 4 to 5\,K. 
\tabularnewline

\begin{acknowledgments}
The authors thank Hubertus Luetkens and Vladimir Pomjakushin for fruitful discussions. This work is partly based on experiments performed at the Institut Laue-Langevin (ILL), Grenoble, France. Neutron data collection (https://doi.org/10.5291/ILL-DATA.5-31-2660) using D11 and D1B at ILL took place with support from proposal 5-31-2660. This work is based on experiments performed at the Swiss Muon Source S\textgreek{m}S, Paul Scherrer Institute, Villigen, Switzerland. Part of the magnetic measurements were carried out on the PPMS/MPMS devices of the Laboratory for Multiscale Materials Experiments, Paul Scherrer Institute, Villigen, Switzerland, as well as the PPMS device of the Kristall- und Materiallabor, Goethe University Frankfurt am Main, Germany. V.U. and J.S.W. acknowledge financial support from the SNF Sinergia CRSII5\_171003 NanoSkyrmionics and SNF 200021\_188707.
\end{acknowledgments}

\bibliographystyle{apsrev4-1}
\addcontentsline{toc}{section}{\refname}\nocite{*}
\bibliography{weyllibrary}

\end{document}